\documentclass[graybox,natbib,nosecnum]{svmult}
\bibpunct{(}{)}{;}{a}{}{,} 

\pdfoutput=1   

\usepackage{mathptmx}       
\usepackage{helvet}         
\usepackage{courier}        
\usepackage{type1cm}        

\usepackage{makeidx}         
\usepackage{graphicx}        
\usepackage{multicol}        
\usepackage[bottom]{footmisc}
\usepackage[normalem]{ulem}	
\usepackage{hyperref}  

\usepackage{soul}   

\newcommand*\aap{A\&A}

\newcommand*\aj{AJ}

\newcommand*\apj{ApJ}
\newcommand*\apjl{ApJ}

\newcommand*\apjs{ApJS}

\newcommand*\gca{Geochim Cosmochim Acta}

\newcommand*\icarus{Icarus}

\newcommand*\mnras{MNRAS}

\newcommand*\nat{Nature}

\newcommand*\pasp{PASP}

\newcommand*\planss{Planet Space Sci}

\newcommand*\procspie{Proc SPIE}

\newcommand{\project}[1]{\textsl{#1}}
\newcommand{\JWST}{\project{JWST}}
\newcommand{\HST}{\project{HST}}
\newcommand{\TESS}{\project{TESS}}
\newcommand{\Spitzer}{\project{Spitzer}}
\newcommand{\Kepler}{\project{Kepler}}

\makeindex             

\begin{document}

\title*{Exoplanet Atmosphere Measurements from Transmission Spectroscopy and other Planet-Star Combined Light Observations}
\titlerunning{Atmosphere Measurements from Combined Light Observations} 
\author{Laura Kreidberg}
\institute{Laura Kreidberg \at Harvard Society of Fellows, 78 Mount Auburn Street, Cambridge, MA, 02138, \email{laura.kreidberg@cfa.harvard.edu}}
%
%
\maketitle

\abstract{It is possible to learn a great deal about exoplanet atmospheres even when we cannot spatially resolve the planets from their host stars. In this chapter, we overview the basic techniques used to characterize transiting exoplanets -- transmission spectroscopy, emission and reflection spectroscopy, and full-orbit phase curve observations. We discuss practical considerations, including current and future observing facilities and best practices for measuring precise spectra. We also highlight major observational results on the chemistry, climate, and cloud properties of exoplanets.}

\section{The Host Star: Friend or Foe?}
Exoplanet atmospheres are a treasure trove of information. By measuring the chemical composition and thermal structure of an atmosphere, it is possible to constrain the planet's formation and evolutionary history, current climate, and even habitability.  Making good on this opportunity is extremely challenging on a technical level, however. In most cases, it is not possible to spatially resolve the planet from its host star, and the star outshines the planet by a factor of at least a thousand to one.  

A number of creative strategies have been developed to circumvent these challenges and make use of the star as a constant reference point.  These approaches are mainly applicable to transiting planets, which periodically pass in front of and behind their host stars. During a transit, a small fraction of the stellar light filters through the planet's atmosphere on its path to us. When the planet is eclipsed, its thermal emission and reflection are blocked.  With meticulous observations, these effects can be measured relative to the constant baseline flux from the star. The resulting inferences of planetary atmospheric properties can be remarkably detailed, given that they were based solely on point sources of light hundreds of parsecs distant.  This chapter is a review of combined light observing techniques, best practices, and science highlights. 


\section{Observing Techniques} 

\subsection{Transit Spectroscopy}
\begin{figure}
\begin{centering}
\includegraphics[scale=.4]{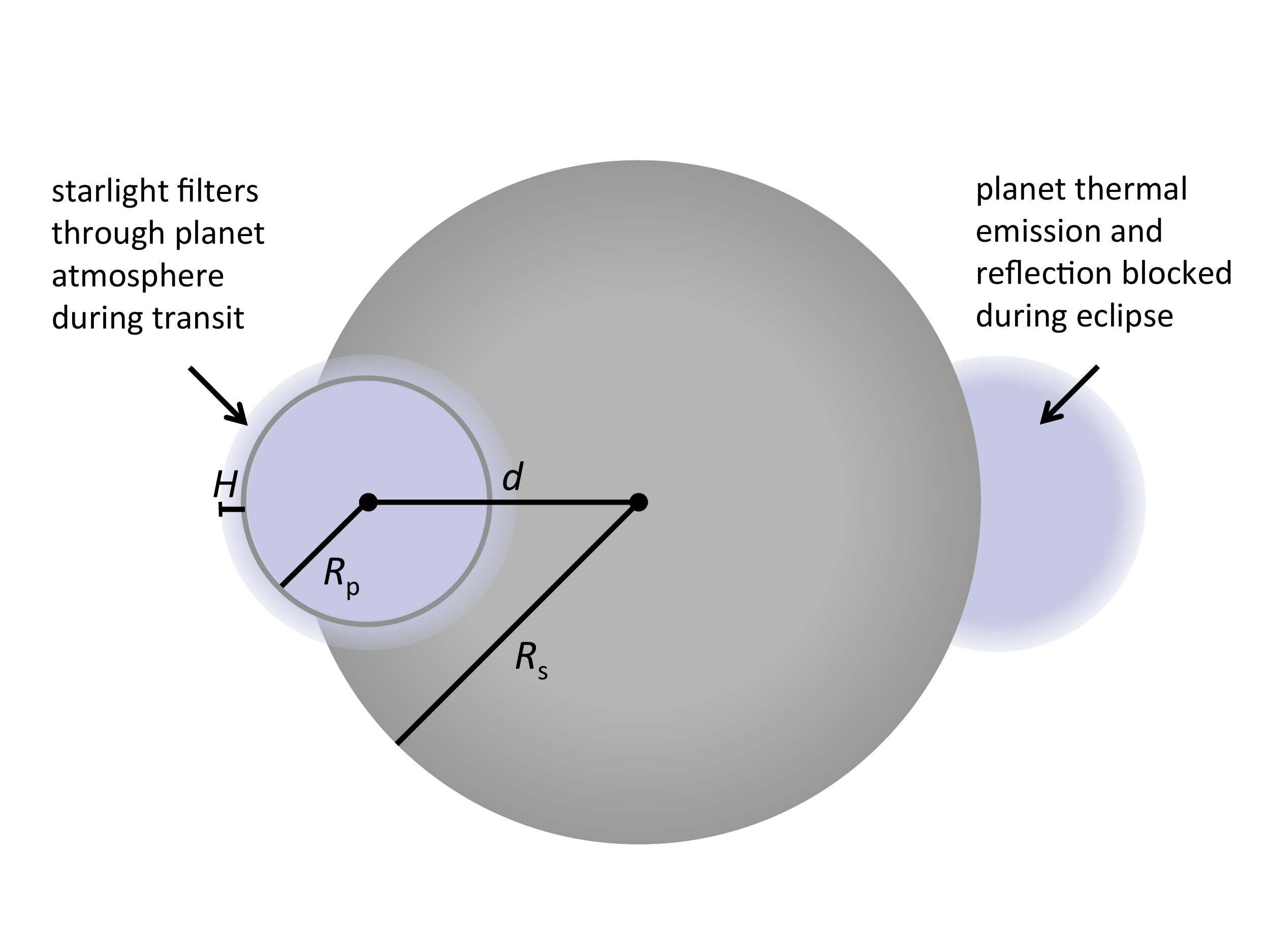}
\caption{Illustration of transit and eclipse geometry. Over the course of the planet's orbit, it periodically passes in front of the star (a transit event) and behind (a secondary eclipse). A few key distances are labeled: the planet and stellar radii, $R_p$ and $R_s$, the atmospheric scale height $H$, and the separation of centers in the plane of the sky, $d$. Figure adapted from \citealt{robinson17} and used with permission from the AAS.}
\label{fig:geom}       
\end{centering}
\end{figure}

\begin{figure}
\begin{centering}
\includegraphics[scale=.8]{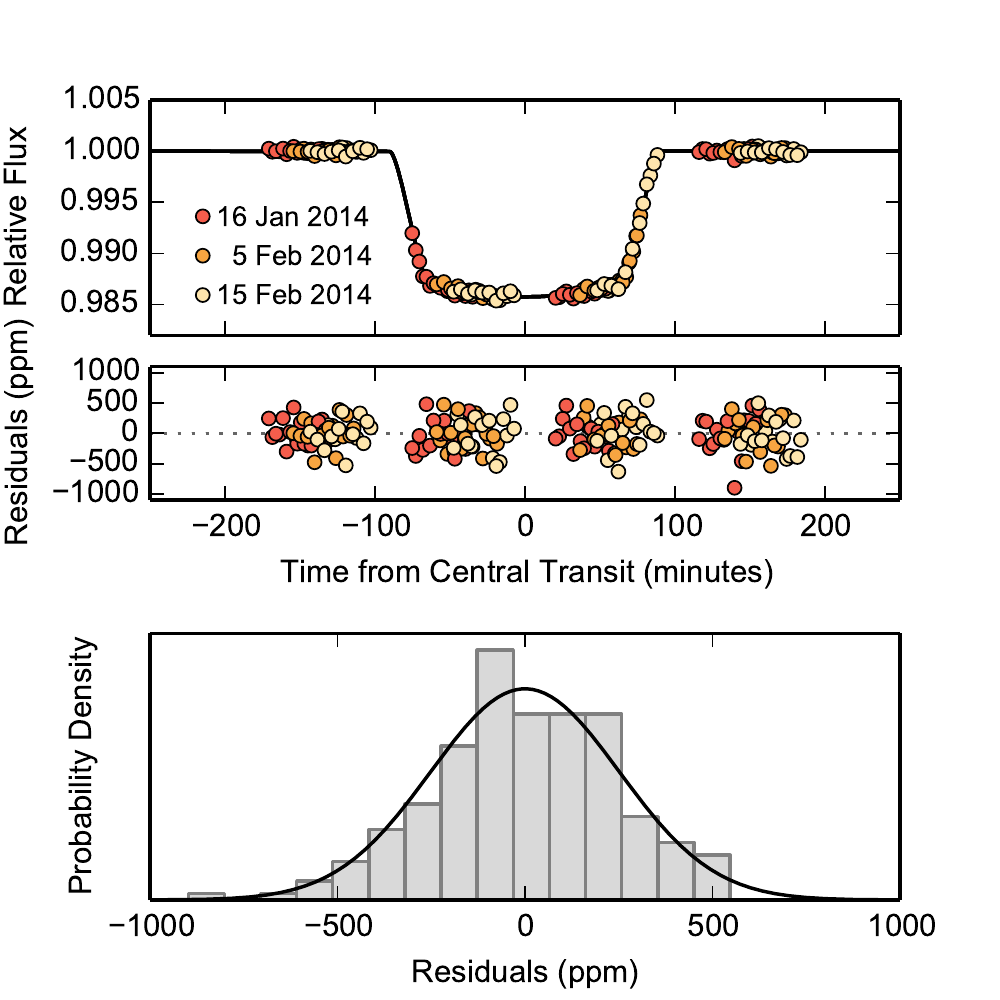}
\caption{Transit light curve measured with the Hubble Space Telescope for the hot Jupiter WASP-12b from \citealt{kreidberg15b} (reproduced by permission of the AAS). The top panel shows the data (points), corrected for instrument systematics and color-coded by observation date, compared to the best fit transit model (line). Gaps in coverage are due to Earth occultation. The middle panel shows the residuals from the best fit, and the bottom panel show a histogram of the residuals compared to the expected photon noise. The light curves are consistent from visit to visit, and the residuals are within 5\% of the photon noise limit, enabling robust estimates of the transit depth and its uncertainty.}
\label{fig:lc}       
\end{centering}
\end{figure}

The most widely used combined-light technique is transmission spectroscopy. For this method, the planet is observed in transit as it passes in front of its host star, as illustrated in Figure\,\ref{fig:geom}.  The measurement of total brightness (star plus planet) over time is known as the transit light curve.  During the transit, the planet blocks a small fraction of the stellar flux equal to the sky-projected area of the planet relative to the area of the star. We refer to this fractional drop in flux as the transit depth, $\delta$. See Figure\,\ref{fig:lc} for an example of a transit light curve.

\begin{figure}
\begin{centering}
\includegraphics[scale=.8]{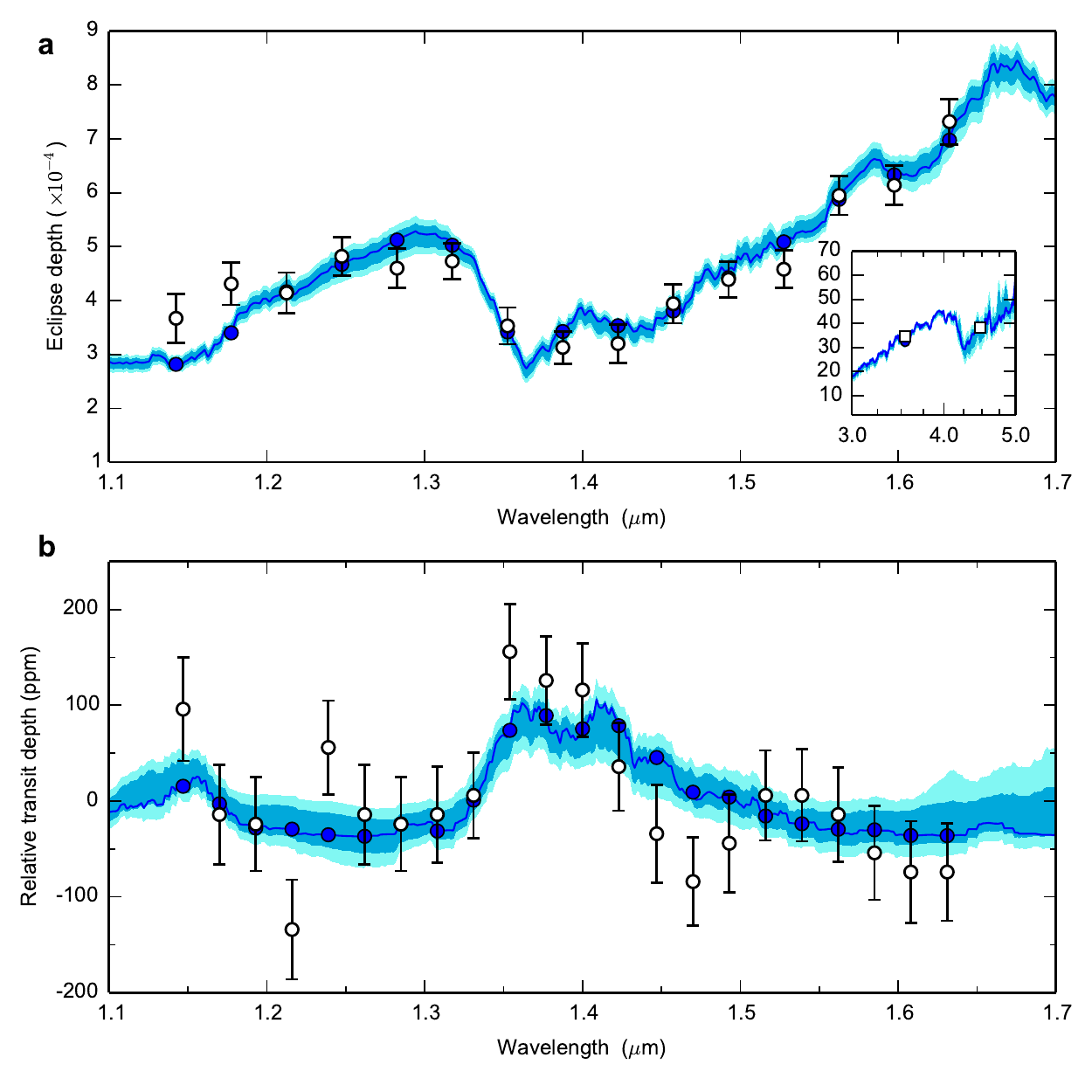}
\caption{Thermal emission spectrum (top) and transmission spectrum (bottom) for the hot Jupiter WASP-43b, compared to best fit models (from \citealt{kreidberg15b}; reproduced with permission from the AAS). The data are from the Hubble Space Telescope WFC3 instrument (1.1 - 1.7 $\mu$m) and Spitzer/IRAC (3.6 and 4.5 $\mu$m). The blue line corresponds to the best fit model, and dark and light blue shading correspond to the 1- and 2-$\sigma$ confidence intervals from an atmospheric retrieval. The retrieval analysis explores the parameter space of temperature profiles and chemical compositions that match the data. In this case, water absorption -- the broad feature at 1.4 $\mu$m -- is detected at high confidence in both spectra ($>5\,\sigma$). This feature is made up of $\sim10^4$ rotational and vibrational line transitions, that are averaged to form the broad-band absorption feature seen here at low spectral resolution. The retrieval constrains the water abundance to be between $0.4$ and $3.5\times$ solar at $1\,\sigma$ confidence.} 
\label{fig:spectra}       
\end{centering}
\end{figure}

The key idea behind transmission spectroscopy is that the planet's transit depth is \emph{wavelength-dependent}.  At wavelengths where the atmosphere is more opaque due to absorption by atoms or molecules, the planet blocks slightly more stellar flux.  To measure these variations, the light curve is binned in wavelength into spectrophotometric channels, and the light curve from each channel is fit separately with a transit model.  The measured transit depths as a function of wavelength constitute the transmission spectrum, so named because the variation arises from the transmission of stellar flux through the planet's atmosphere. Figure\,\ref{fig:spectra} shows the near-infrared transmission spectrum of the hot Jupiter WASP-43b, which has a strong water absorption feature centered at 1.4 $\mu$m. 
 
Theoretical models for the transmission spectrum require radiative transfer calculation for light on the slant path through the planet's atmosphere \citep{seager00}, a computationally intensive task. However, we can make a rough prediction of the size of features in the transmission spectrum based on the atmospheric scale height $H$. The scale height is the change in altitude over which the pressure drops by a factor of $e$. Assuming hydrostatic equilibrium and using the ideal gas law,

\begin{equation}
H = \frac{K_bT_{eq}}{\mu g}
\end{equation}
where $K_b$ is the Boltzmann constant, $T_{eq}$ is the planet's equilibrium temperature, $\mu$ is the mean molecular mass, and $g$ is the surface gravity.

The amplitude of spectral features in transmission is then:
\begin{eqnarray}
\delta_\lambda &=& \frac{(R_p + nH)^2}{R_s^2} - \frac{R_p^2}{R_s^2}\\
 & \approx & 2nR_pH/R_s^2
\end{eqnarray} 
where $n$ is the number of scale heights crossed at wavelengths with high opacity (typically around two for cloud-free atmospheres at low spectral resolution; \citealt{stevenson16}). It follows that ideal candidates for transmission spectroscopy have high equilibrium temperatures, small host stars, low surface gravity, and low mean molecular mass composition (hydrogen-dominated). But even for these ideal cases, the amplitude of spectral features is just $\delta_\lambda \sim0.1\%$. For Earth-like planets, the expected amplitude is two to three orders of magnitude smaller, depending on host star size. 

\subsection{Occultation Spectroscopy}
A corollary of the transit spectroscopy method is occultation spectroscopy, which measures thermal emission and reflection from the planet. Rather than observing the planet during transit, it is observed at secondary eclipse, when it passes behind the host star. The eclipse provides a baseline measurement of the stellar flux alone. When the planet orbits back into view, any increase in brightness can be attributed to the planet's thermal emission and reflected light. Secondary eclipses are discussed in detail in the chapter ``Characterization of exoplanets: Secondary eclipses" by Alonso.

\runinhead{Thermal Emission}
For the short-period planets that have been studied so far, the dominant source of thermal emission is re-radiation of incident stellar flux (rather than latent heat of formation, as seen for directly imaged planets). Thus the typical size of the emission signal can be predicted from the planet's equilibrium temperature:

\begin{equation}
\label{eqn:fpfs}
\frac{F_p}{F_s} = \frac{B(\lambda, T_{eq})}{B(\lambda, T_s)}\left(\frac{R_p}{R_s}\right)^2
\end{equation}
where $F_p/F_s$ is the planet-to-star flux ratio, $B(\lambda, T)$ is the blackbody spectral radiance at temperature $T$, and $R_p/R_s$ is the planet-to-star radius ratio. Since the planet is cooler than the star, the flux ratio is larger at longer wavelengths. For example, the planet-to-star flux for the hot Jupiter HD 209458b is just 50 parts per million at 1 $\mu$m, but increases to over 1000 ppm at 4.5 $\mu$m \citep{line16}.

Equation \ref{eqn:fpfs} is a good first order approximation of the planet signal, but as with transmission spectroscopy, more complex features arise in the spectrum due to the atmosphere's changing opacity with wavelength. The emitted light comes from the photosphere of the planet, where the optical depth is of order unity. At more opaque wavelengths, the photosphere is at higher altitude, where the temperature may differ from $T_\mathrm{eq}$. We illustrate this effect in Figure\,\ref{fig:spectra} with the emission spectrum of WASP-43b, which has a strong water absorption feature. In the water band at centered at 1.4 $\mu$m, we see a higher, cooler layer of the planet's atmosphere, where the planet's flux is lower.

The size and shape of spectral features depends on the exact temperature-pressure profile of the atmosphere -- for example, if the temperature increases with altitude (known as a thermal inversion or stratosphere) spectral features can be seen in emission rather than absorption. Thermal emission spectroscopy is therefore a useful probe of temperature structure in addition to atmospheric composition.


\runinhead{Reflected Light}
At short wavelengths, reflected light from the planet may also be detectable. It is convenient to quantify the amount of reflected light in terms of a ``perfect" Lambertian surface: a flat, perfectly diffusing disk with the same cross-sectional area as the planet. The ratio of reflected light from the fully illuminated planet, relative to reflection by a perfect mirror, is the geometric albedo $A_g$. The total reflected light signal is

\begin{equation}
F_{reflect} = A_g(R_p/a)^2 \Phi(\alpha)
\end{equation}
where $a$ is the orbital separation and $\Phi(\alpha)$ is the phase function (the reflected light intensity at phase angle $\alpha$). The phase function depends on the scattering properties of the atmosphere, but analytic predictions are available for certain simplified models \citep[e.g.][]{madhu12}.  Reflected light is easiest to detect in the optical, where it dominates the thermal emission signal; however, the amplitude tends to be small \citep[typically less than 100 ppm;][]{angerhausen15}.

\runinhead{Phase Curves}
In addition to characterizing thermal emission and reflected light from secondary eclipses, it is also possible to observe a full-orbit phase curve \citep{seager00b, knutson07}. These observations consists of continuous time series photometry or spectroscopy of a planet over its entire orbital period, using the secondary eclipse as a baseline measurement of the stellar flux alone.  The targets for phase curve observations so far have been short-period planets that are tidally locked to their host stars. For these planets, the rotation period is known (equal to their orbital period), so over the course of one complete orbit all longitudes are visible in turn.  Phase curves are powerful because they probe atmospheric physics and chemistry over a wider geographic region than transits or eclipses. The theory and observations of phase curves are discussed further in the chapter ``Characterization of exoplanets: Observations and modeling of orbital phase curves" by Parmentier \& Crossfield.

\runinhead{Eclipse Mapping}
All of the occultation spectroscopy techniques described above rely on a hemisphere average of the planet's reflected or emitted light. To glean additional spatial information, one can use the eclipse mapping method, which requires precise observations of the secondary eclipse light curve during ingress and egress.  During these intervals, the hemisphere of the planet is partially eclipsed by the star, allowing the observer to pinpoint the brightness of a smaller region and thus map the brightness distribution in detail \citep{rauscher07, dewit12}.  This is the only combined light technique that is sensitive to the planet's brightness as a function of \emph{latitude} (because higher latitudes are first to enter and exit eclipse). For additional discussion of mapping techniques, see the chapter ``Mapping of exoplanets" by Cowan \& Fujii.

\section{Practical Considerations}
Detecting the tiny signals from exoplanet atmospheres is a formidable challenge. Even for the most favorable systems, the amplitude of spectral features is of order a tenth of a percent. Pushing to this level of precision requires many photons (and thus large telescopes), a stable observing environment, and detailed knowledge of the planet's host star.  In this section, we summarize the state-of-the-art in observing facilities, briefly review ground-based observing techniques, and discuss possible sources of error.

\subsection{Observing Facilities}
The Hubble and Spitzer Space Telescopes are the preeminent facilities for exoplanet atmosphere studies.  Space-based observations have several advantages: they are free from atmospheric turbulence that adds systematic noise to light curves, they can access wavelengths where the Earth's atmosphere is strongly absorbing (e.g., ultra-violet and water absorption bands), and they have lower thermal background noise in the infrared, facilitating long-wavelength observations. 

\begin{figure}
\begin{centering}
\includegraphics[width=\textwidth, trim={0.5cm 0cm 0cm 0.1cm},clip]{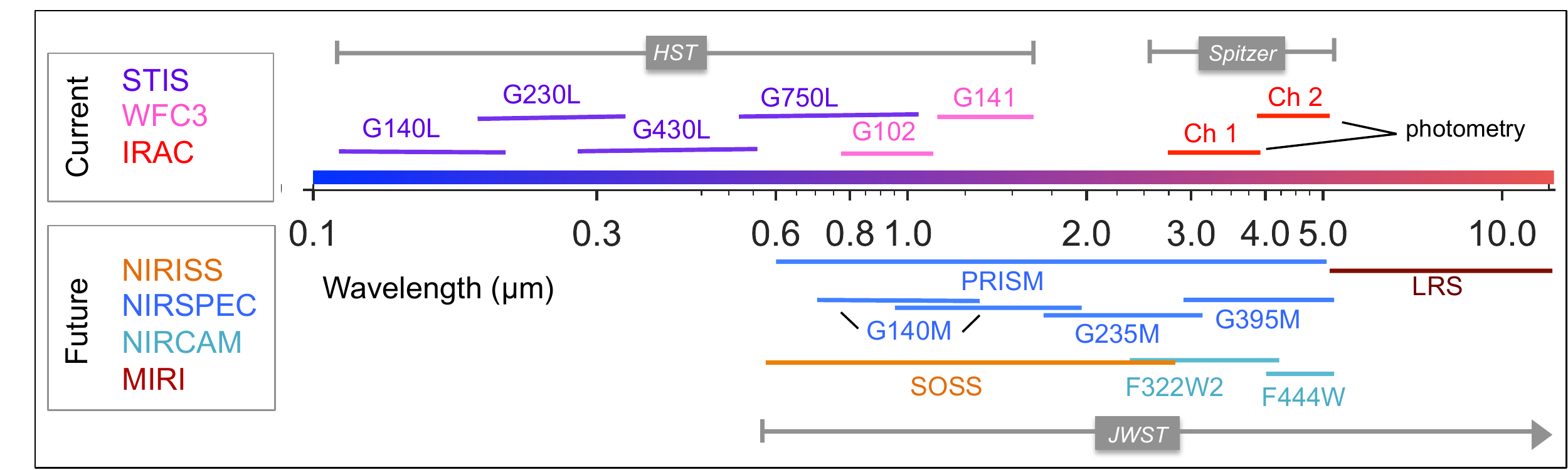}
\caption{Summary of current (circa 2017) and planned observing capabilities from space. The color bar in the center indicates the wavelength scale in microns. Current instrumentation covers the UV/optical (HST/STIS), near-infrared (HST/WFC3) and infrared (Spitzer/IRAC photometry). The four instruments on board JWST (NIRISS, NIRCAM, NIRSPEC, and MIRI) will provide spectroscopy from 0.6 to 12 $\mu$m. The typical resolution is $\lambda/\Delta \lambda \sim100 - 1000$. Not shown are HST and JWST photometric filters, or higher-resolution dispersive elements.}
\label{fig:instruments}       
\end{centering}
\end{figure}

As of 2017, the workhorse instruments for atmosphere characterization are the Space Telescope Imaging Spectrograph (STIS) and the Wide Field Camera 3 (WFC3) on board Hubble, and Spitzer's Infrared Array Camera (IRAC).  These instruments provide spectroscopy from the UV to the near-infrared (ending at 1.7 $\mu$m), and broadband photometry at 3.6 and 4.5 $\mu$m. The suite of instruments built for the James Webb Space Telescope (JWST; scheduled for launch in 2018), will provide significantly expanded spectroscopic coverage, from $0.6$ to $12\,\mu$m.  Figure\,\ref{fig:instruments} illustrates the observing capabilities of these current and planned instruments. 

\subsection{Ground-Based Observations}
Exoplanet atmosphere characterization is possible from the ground, but it is challenging because
the transparency of Earth's atmosphere changes with time.  As a target moves through the sky, line-of-sight properties such as the airmass, precipitable water vapor, and cloud coverage will vary, introducing systematic trends in the light curve that may be orders of magnitude larger than the signal from the planet. 

To correct for these effects, one approach is to use comparison stars to correct for systematic trends in the target light curve. In addition to the target, several nearby stars are observed in the same field of view. The path of their flux traverses similar parts of the Earth's atmosphere, so their light curves exhibit nearly identical systematic trends as the target. The target light curve can then be divided by the sum of the comparison star light curves to remove systematics.  For spectroscopic observations, the target and comparison stars must be observed with a custom-made mask with wide slits ($>10''$) to avoid time-dependent slit losses due to variable seeing, a technique developed by \cite{bean10}.

Another creative ground-based technique uses high resolution spectroscopy to detect the planet's atmosphere.  The idea is that the planet spectrum is Doppler shifted due to its orbital motion, and thus separated in wavelength from both the stellar spectrum and telluric lines from the Earth's atmosphere. The planet spectrum can be cross-correlated with a template to reveal the atmospheric composition and orbital velocity \citep[e.g.][]{snellen10}.  This technique is discussed in detail in the chapter ``Spectroscopic direct detection of exoplanets" by Birkby. 

\subsection{Noise Sources}
\runinhead{Photon noise} 
Stars emit $N \pm \sqrt{N}$ photons per unit time. The $\sqrt{N}$ noise, so-called photon noise, arises because each atom in the star emits a photon with some small probability per unit time, so the count rate follows a Poisson distribution. Photon noise is the fundamental limit on the precision of a light curve -- the only way to improve the precision on a planet's spectrum is to stack many observations together. Bright stars have lower photon noise than faint stars (the relative error $\sqrt{N}/N$ decreases as $N$ increases), so the best target systems for atmosphere characterization typically have H mag $< 10$.  Space-based observations often reach the photon limit \citep{sing11, deming13, ingalls16}, whereas ground-based observations are typically a factor of a few above it \citep[e.g.][]{bean13}. 

\runinhead{Instrument Systematics}
Instrument-based systematics are a major source of error in precise time series photometry, and can be orders of magnitude larger than the planet signal. Three common systematic effects are: 

\begin{itemize}
\item{\emph{Charge trapping}. Near-infrared detectors (e.g. \HST/WFC3) have impurities that can trap photoelectrons \citep{smith08}. As the traps fill up, the number of recorded photoelectrons increases exponentially, with a time constant that depends on the detector illumination history and current count rate \citep{zhou17}.  This same mechanism is responsible for image persistence, the afterglow that appears as charge traps are released. The effect can be corrected with analytic models to photon-limited precision \citep{deming13, line16}. 
}
\item{\emph{Intrapixel effect.} Detector pixels do not have perfectly uniform spatial sensitivity, so the measured flux is correlated with the $X, Y$ position of the centroid of the image. This effect is the dominant systematic error for infrared measurements (e.g. \Spitzer/IRAC). Many techniques have been developed to model intrapixel variation, including \texttt{BLISS} mapping \citep{stevenson12}, pixel-level decorrelation \citep{deming15}, and ICA \citep{morello15}.}
\item{\emph{Variable illumination.} The position of the spectrum on the detector can shift slightly over the course of an observation, either due to pointing drift or changes in telescope focus. These shifts cause light to fall on pixels that may have different sensitivity, and the flatfield correction is generally not known well enough to remove the induced variation. This effect can sometimes be corrected with a polynomial fit \citep[e.g. for HST/STIS observations;][]{sing11}, but in other cases can produce a large noise floor of order $\sim0.1\%$ \citep[as for the old HST instrument NICMOS;][]{gibson11}.}
\end{itemize}

\runinhead{Astrophysical Systematics}
Systematic errors also arise from incorrect or incomplete models for the stellar or planet flux. These include: 

\begin{itemize}
\item{\emph{Background stars.} Roughly half of stars have one or more bound companions \citep{raghavan10}. If the companion flux is blended with that of the host star, it dilutes the planet signal. Transit and eclipse depths must be multiplied by a correction factor $1 + \beta(\lambda)$ where $\beta$ is the ratio of the background star to host star flux. High contrast imaging is needed to detect close companions, and should be obtained for systems that are targets for atmosphere characterization.}
\item{\emph{Unocculted star spots.} To first order, the temperature difference between the stellar photosphere and the spotted region introduces a slope in the planet spectrum. If the spots are cool enough for molecules to form (e.g. water), they can even produce spurious spectral features, as there is no way to distinguish between absorption due to molecules in the spot and absorption by the planet atmosphere. The effect from spots can be corrected by multiplying the planet spectrum by a factor $(1-s\times[1-F_{\lambda,\mathrm{spot}}/F_{\lambda,\mathrm{phot}}])^{-1}$, where $s$ is the spot covering fraction and $F_\lambda$ is spectral radiance for the spot or the photosphere \citep{mccullough14}. The spot properties can be estimated based on the amplitude of long-term photometric variability in the stellar lightcurve or from spot crossing during transit \citep[e.g.][]{pont08}.} 
\item{\emph{Stellar activity.} Variations in star spot coverage are also a source of bias in transit depth measurements. To correct transit depths taken at different epochs, one can obtain photometric monitoring of the host star to estimate changes in $s$ and correct the depths with the above scale factor.} 
\item{\emph{Nightside emission from the planet.} For the hottest planets, thermal emission from the nightside may contribute significant flux during the transit. The nightside temperature depends on the planet's heat redistribution and cloud coverage, so can only be measured directly from phase curve observations. To correct for nightside flux, the transit depths should be multiplied be a factor $(1 + F_p/F_s)^{-1}$, where $F_p$ is the planet nightside flux and $F_s$ is the stellar flux \citep{kipping10}.} 
\end{itemize}

\runinhead{Notes on Reliability}
We have noted here that there are many potential challenges for precise atmosphere characterization, and indeed, several early results were later disputed \citep[e.g.][]{tinetti07, swain08, gibson11}. More recently, however, observing techniques and data analysis have advanced to the point that space-based measurements are generally photon noise-limited, repeatable, and reproducible by multiple teams \citep[e.g.][]{deming13, kreidberg14a, ingalls16}. The best achieved precision to date is 15 ppm on the emission spectrum of HD 209458b at a resolution $R\sim 30$ \citep{line16}, based on five secondary eclipse observations. These results bode well for the precision achievable with JWST, which will have newer detectors and better pointing stability. 

Ground-based observations have not been as thoroughly vetted as space-based data.  The comparison star technique can yield light curves that are within a factor of a few times the photon noise limit, but it can be difficult to estimate uncertainties correctly in the presence of residual correlated noise \citep[e.g.][]{jordan13, beatty16}.  Pushing the performance of ground-based telescopes will be important in coming years, as they will be the only facilities capable of observing wavelengths shorter than $0.6\,\mu$m after Hubble is no longer operating. 

\section{Major Results from Atmosphere Studies}
In this section we discuss observational highlights, with a focus on space-based transmission and emission spectroscopy.  For recent comprehensive reviews, see \cite{crossfield15} and \cite{deming17}. 

\subsection{Atmospheric Composition}
\runinhead{Expectations}
We know from the Solar System that atmospheric composition can vary widely from planet to planet. Even so, we can still make educated guesses about possible compositions based on the building blocks of planet formation,  which come from disks of gas and dust surrounding young stars.  The protoplanetary disk has similar composition to the host star: predominantly hydrogen and helium, with smaller amounts of metals (the most abundant being oxygen, carbon, and nitrogen; \citealt{anders89}).  

\begin{figure}
\begin{centering}
\includegraphics[width=\textwidth]{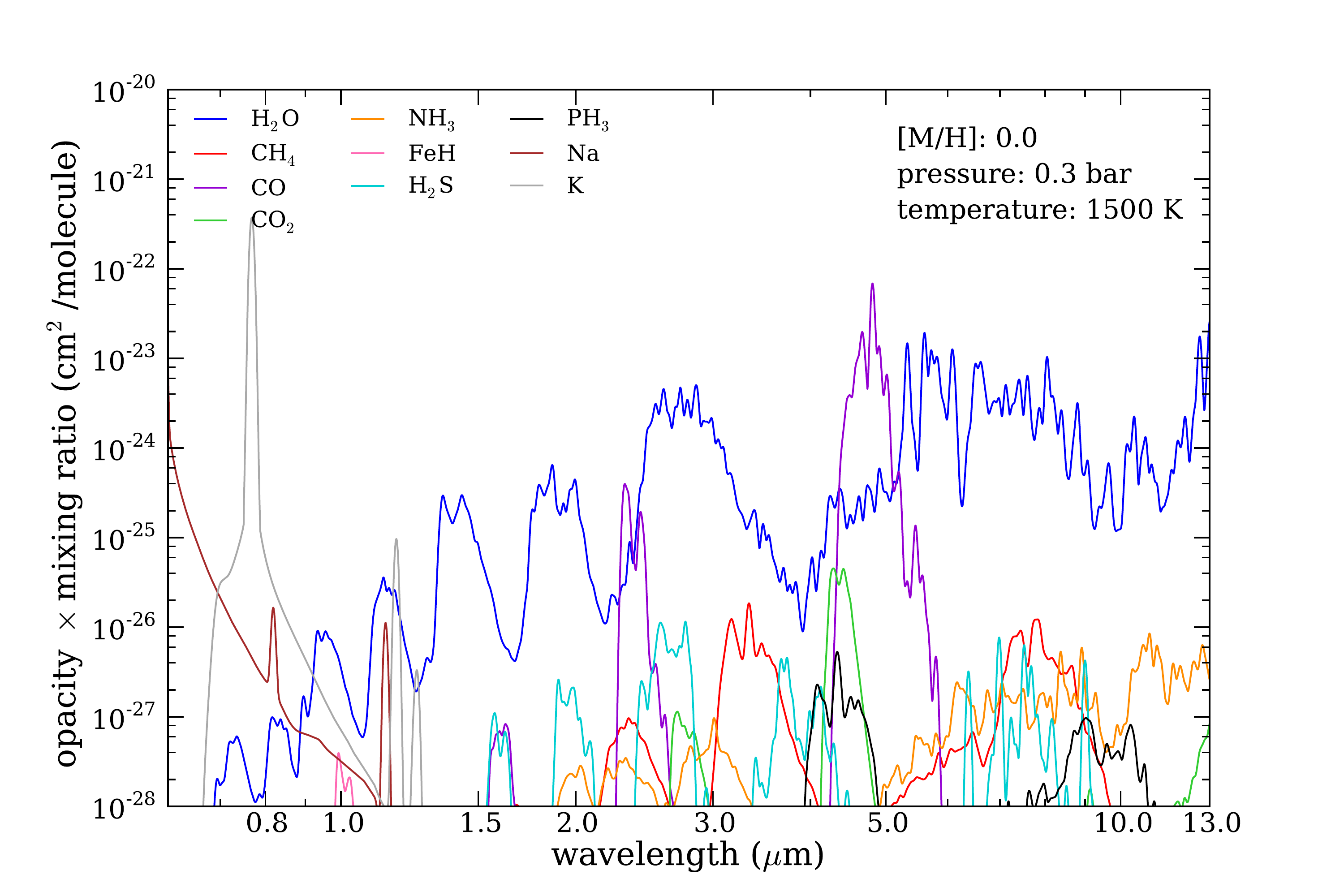}
\caption{Opacities for a solar composition atmosphere in chemical equilbrium at 1500 K and 0.3 bar. Sodium and potassium are the dominant absorbing species at optical wavelengths, whereas water and carbon monoxide are dominant in the near-infrared. At lower atmospheric temperatures (not shown), methane and ammonia become more abundant.  Even though H and He dominate the atmospheric composition, they only have collision-induced continuum opacity at infrared wavelengths, and the spectrum is instead dominated by the molecular and atomic species shown here.  Figure made with the Opacity Wizard tool, courtesy of Caroline Morley (https://github.com/astrocaroline/opacity-wizard).}
\label{fig:opacity}       
\end{centering}
\end{figure}

These main constituent elements are expected to combine into H$_2$, H$_2$O, CO, CO$_2$, CH$_4$, NH$_3$, O$_2$, and N$_2$, depending on the temperature, pressure, and composition of the planet's atmosphere \citep{moses13}.  Many of these species are easily detectable due to their strong absorption features, as illustrated in Figure\,\ref{fig:opacity}.  This figure shows the predicted opacity of dominant absorbers for a solar composition atmosphere at 1500 K (representative of a typical hot Jupiter).  In addition, there are several species that are less abundant but still detectable thanks to their large absorption cross sections: namely, the alkali metals (sodium and potassium) and titanium and vanadium oxides (TiO and VO). 

\runinhead{Detections}
Many of the predictions listed above have been borne out.  The first detection of an absorbing species was the sodium resonance doublet at 589 nm in the hot Jupiter HD 209458b \citep{charbonneau02}. Since then, both sodium and potassium have been detected in a number of other hot Jupiters \citep[e.g.][]{sing11b, nikolov14}. 

Hydrogen has also been detected.  Lyman-$\alpha$ observations with HST/STIS revealed large exospheres of escaping atomic hydrogen around hot planets \citep[e.g.][]{vidal-madjar03, ehrenreich15}.  These exospheres also contain ionized metals \citep[e.g., OI, CII;][]{vidal-madjar04}. For more on atmospheric escape, see the chapter by Lecavelier and Bourrier, ``Characterizing evaporating atmospheres of giant exoplanets". In addition to atomic hydrogen, the presence of molecular H$_2$ was inferred from the Rayleigh scattering signature in the spectrum of HD 209458b \citep{lecavelier08}. However, condensates can also cause Rayleigh scattering, so it is not certain that H$_2$ is the culprit. 

The most commonly observed molecular species is water, which has been seen in over a dozen planets \citep[e.g.][]{deming13, huitson13, birkby13, mccullough14, fraine14, kreidberg14b, kreidberg15b, line16, evans16, stevenson16b}. Most of these results come from HST/WFC3 spectra in the near-IR (1.1 - 1.7 $\mu$m). WFC3 observations have provided the most successful detections of molecular features, thanks to the combination of the instrument's state-of-the-art measurement precision and water's strong absorption features in the near-infrared.  To illustrate, Figure\,\ref{fig:spectra} shows an example of water features in WFC3 transmission and emission spectra for WASP-43b \citep{kreidberg14b}.  


Carbon-bearing molecules have proven harder to detect, mainly due to the wavelength range accessible by current observing facilities. Hot Jupiters are expected to have strong CO features in the near-infrared (see Figure\,\ref{fig:opacity}), but the \Spitzer\ photometric bandpasses are about $1\,\mu$m wide and cover spectral features from multiple different absorbing species (see Figure \ref{fig:instruments}). Several results from Spitzer suggest absorption from carbon-bearing species, but it is challenging to unambiguously identify which absorber is present \citep[e.g.][]{desert09, stevenson10, madhusudhan11, morley17}.  The only definitive detections of CO are from high-resolution ground-based spectroscopy \citep{dekok13,brogi14}.  

There are some noteworthy molecules that have not yet been seen. These include ammonia and methane, which are abundant in the atmospheres of the Solar System gas giants. Their absence in exoplanet transmission spectra is not a big surprise: these molecules are unstable at high temperatures, and will more likely be seen in cooler targets than those observed so far.  Despite many searches, there is no definitive evidence for TiO/VO either \citep[e.g.,][]{sing13, evans16}. These molecules are expected in hot planets \citep{fortney08}, but they may rain out or get cold trapped deep in the atmosphere \citep{parmentier13}.

We note that nearly all of the results listed here are for hot Jupiters, but there have been a few detections for smaller, cooler planets. Water features were inferred in the transmission spectra of two exo-Neptunes with equilibium temperatures below 1100 K \citep{fraine14, wakeford17}. The thermal emission spectrum of the warm Neptune GJ 436b also shows a strong absorption feature in the Spitzer 4.5 $\mu$m filter, which could be due to CO/CO$_2$ \citep{morley17}.  These observations have only been possible for a handful of the most observationally feasible small planets, but many more systems will become accessible with JWST thanks to its larger collecting area. 

\runinhead{Absolute Abundances}
Once an atom or molecule has been detected, a next step is to retrieve its abundance (as described in the chapter ``Atmospheric Retrieval for Exoplanet Atmospheres" by Madhusudhan). Abundance measurements are useful because they are diagnostic of planetary formation conditions, such as the surface density of solids and the relative accretion rates of gas and ice \citep[e.g.][]{fortney13, mordasini16}. Abundances are typically quantified in terms of their enrichment over the solar value (i.e., the expected abundance for a solar composition gas at the temperature of the planet's atmosphere).  The more precise the measurement, the better it can constrain the planet's formation history. However, precise determinations are challenging, even if molecular absorption is detected at high confidence.  Based on transmission spectra alone, there are order of magnitude degeneracies between chemical mixing ratios and atmospheric pressure \citep{benneke12, griffith13}.  

\begin{figure}
\begin{centering}
\includegraphics[scale=.8]{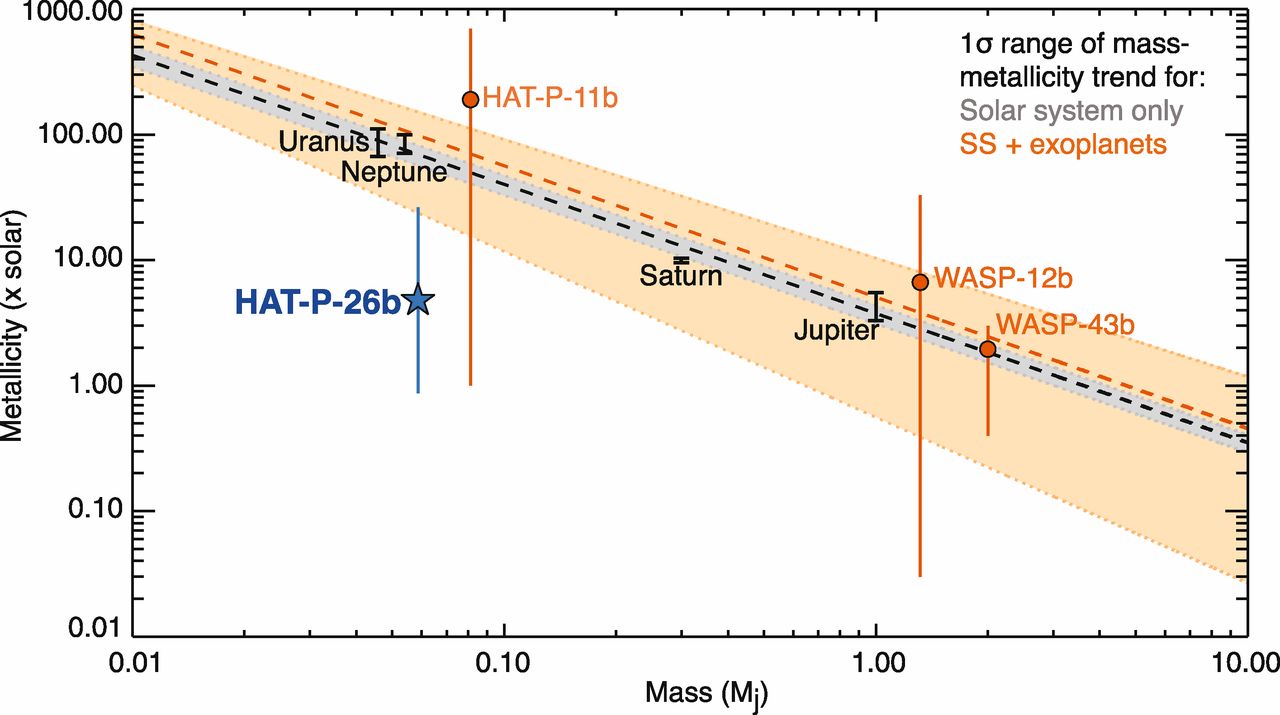}
\caption{Atmospheric metallicity versus planet mass for the Solar System planets and exoplanets (from \citealt{wakeford17}; reprinted with permission from AAAS.). The Solar System shows a pattern of decreasing metallicity with increasing planet mass, which is fit well by a power law (black dashed line with gray shaded 1-$\sigma$ uncertainties). Exoplanets roughly hold to this pattern, but may exhibit more scatter (orange line with shading). While exoplanet atmospheric metallicities have larger uncertainties than for the Solar System planets, it is possible to measure abundances for hundreds to thousands of them, enabling statistical constraints on the metal enrichment pattern.} 
\label{fig:massZ}       
\end{centering}
\end{figure}

To break this degeneracy, one approach is to observe the planet's thermal emission.  Emission spectra are more sensitive to absolute abundance because the shape and amplitude of spectral features depend on the temperature-pressure profile. For example, \cite{stevenson17} measured the water abundance for the hot Jupiter WASP-43b to better than a factor of five  ($0.3 - 1.7\times$ solar at $1\,\sigma$ confidence). Another technique to improve precision requires observe multiple absorption bands for the same molecule in transmission \citep{benneke12}.  \cite{wakeford17} detected two water features in the transmission spectrum of the warm Neptune HAT-P-26b and retrieved a water abundance of $0.8 - 26\times$ solar at 1\,$\sigma$ \citep{wakeford17}. Even though these estimates are much less precise than what has been achieved for the Solar System (see Figure \,\ref{fig:massZ}), there is a much larger and more diverse sample of exoplanets available to study.  A dedicated atmosphere characterization mission could measure abundances for hundreds of planets, and provide a complementary test of planet formation models in addition to detailed study of Solar System planets \citep{chapman17}.

The existing exoplanet metallicity constraints are already an intriguing point of comparison. In the Solar System, there is a pattern of increasing atmospheric metallicity with decreasing planet mass  -- from a factor of a few enhanced over solar for Jupiter to $\sim100\times$ solar for Uranus. This trend is a natural outcome of planet population synthesis models, which show that lower mass planets are relatively more polluted by infalling planetesimals \citep{fortney13, mordasini16}. Using water abundance as a tracer of metallicity, we see that WASP-43b agrees well with the Solar System pattern, but HAT-P-26b is marginally less enriched than expected (as shown in Figure\,\ref{fig:massZ}).  These results hint at a diversity of atmospheric compositions, and provide a proof of concept that precise abundance measurements are possible for exoplanets. It is also exciting that these measurements are for water, specifically. The water abundances for the Solar System gas giants are poorly constrained, because water is condensed deep in their atmospheres \citep{showman98, mousis14}. By contrast, water is well-mixed and in the gas phase for hot exoplanets, providing an opportunity to directly study this critical building block for planet formation.

\runinhead{Abundance Ratios}
The composition of the protoplanetary disk is not uniform everywhere. As the disk temperature drops with radial distance from the star, molecules reach their freezing points, beginning with H$_2$O and followed by CO$_2$ and CO, leading to variation in abundance ratios for gas versus solid material. These gradients in chemistry can be imprinted on the composition of forming planets \citep[e.g.][]{oberg11, madhusudhan14, alidib16}. Because planets are enriched in metals by planetesimal accretion, the abundance pattern is expected to trace that of the solids rather than the gas \citep{mordasini16, espinoza17}.

The carbon-to-oxygen ratio (C/O) has been proposed as a natural framework for characterizing atmospheric abundance patterns, since these are major building blocks for planet formation \citep{madhusudhan12}.  Equilibrium chemical abundances change dramatically when carbon is more abundant than oxygen, so carbon-rich compositions (C/O $> 1$) should be easily identifiable \citep{moses13}.  There was early evidence for a carbon-rich atmospheric composition for WASP-12b based on its thermal emission spectrum \citep{madhusudhan11}, but further scrutiny of this planet and other hot Jupiters has revealed $\mathrm{C/O} < 1$ in all cases \citep{line14, kreidberg15b, benneke15, barstow17}. More precise constraints await JWST, which is expected to constrain C/O for hot Jupiters to a precision of 0.2 dex \citep{greene16}.

\subsection{Climate}
With thermal emission measurements, it is possible to characterize a planet's climate in detail.  To date, these observations have focused on hot Jupiters with large planet-to-star flux ratios. The most common measurement is of dayside brightness temperature, which has been inferred from secondary eclipse observations for over 50 planets \citep{schwartz15}. Measured dayside temperatures range from roughly 1000 to 3000 K \citep{stevenson14b, kammer15, morley17}.  In addition to brightness temperatures, one can also determine the atmosphere's temperature-pressure profile using emission spectroscopy. At wavelengths where the atmosphere has higher opacity, the emitted light comes from higher altitudes. 

\runinhead{Thermal Inversions}
\cite{fortney08} predicted that hot Jupiter atmospheres are divided into two classes: those with thermal inversions (temperature increasing with height) and those with non-inverted profiles. The dividing factor is whether gaseous TiO/VO are present in the atmosphere. These molecules are very strong optical absorbers, and can significantly heat the upper atmosphere even when present in trace amounts. But they are only expected to be present in the hottest atmospheres (flux $>10^9\,\mathrm{erg}\,\mathrm{s}^{-1}\,\mathrm{cm}^{-2}$); at cooler temperatures they will condense and rain out of the atmosphere.

Early \Spitzer\ observations of the hot Jupiter HD 209458b, which falls on the dividing line between the two classes, suggested the atmosphere has a thermal inversion \citep{knutson08}. 
However, new data showed that this was not the case, and that the temperature profile is actually decreasing with height \citep{diamond-lowe14, schwarz15, line16}.  More recently, there has been evidence for weak inversions (isothermal profiles) in the atmospheres of other planets \citep{stevenson14, haynes15}. All of these planets have extremely hot dayside temperatures ($> 2500$ K), suggesting that temperature does play a role in determing thermal structure. 

\runinhead{Global Heat Circulation}
Thermal phase curves provide many additional insights into global climate. This topic is discussed in detail in the chapter by Parmentier and Crossfield, so we will touch on just a few of the highlights. 
\begin{itemize}
\item{The first phase curve measurement, for the tidally locked hot Jupiter HD 189733b with Spitzer by \cite{knutson07}. They measured a small difference in temperature of $250$ K from the dayside to the nightside, and an offset in peak brightness to the east of the substellar point. These results agree well with expectations from 3D global circulation models (GCMs), which predict efficient heat redistribution for slower rotation periods, and eastward equatorial jets caused by the interaction of Rossby waves with the planet's rotation \citep{showman09}.}
\item{The first spectroscopic phase curve, for the hot Jupiter WASP-43b with HST/WFC3 \citep{stevenson14}. The spectra are sensitive to a range of pressure levels in the atmosphere, enabling a determination of the planet's thermal structure as a function of longitude \emph{and} altitude. The planet has a low Bond albedo (0.2), a large day-night temperature contrast (suggesting there are clouds on the nightside), and a hot spot that is shifted further east at higher pressures, indicative of more efficient heat transport deeper in the atmosphere.}
\item{A Spitzer phase curve for the super-Earth (2\,$R_\oplus$) 55 Cancri e \citep{demory16}.  This is the first phase curve measured for a small planet.  The measurement revealed a dayside temperature of 2700 K, with a 1300 K drop in temperature to the nightside. There is a large hot spot offset (40 degrees east of the substellar point).  The results are consistent with either molten rock on the dayside, or an optically thick atmosphere with minimal heat redistribution.}
\end{itemize}

\subsection{Condensates}
Condensates -- an umbrella term that includes clouds and hazes -- are ubiquitous in the Solar System planet atmospheres and are proving to be common in exoplanets as well. Roughly half of hot Jupiters have evidence for condensates in their spectra \citep{sing16}. Transmission spectra are particularly sensitive to the presence of condensates, due to the slant viewing geometry observed during transit \citep{fortney05}.  Condensates have three main effects on transmission spectra.  First, they block transmission of stellar flux, effectively truncating spectral features below the cloud-deck height.  They can also introduce a slope in the spectrum over wavelength intervals of several microns \citep[e.g][]{sing16}.  The third effect is scattering off cloud and haze particles at optical wavelengths, which introduces a steep increase in transit depth towards the blue \citep[e.g.][]{pont08}.

The first hint of extrasolar condensates came from HD 209458b, which had a smaller than expected sodium feature relative to a clear atmosphere \citep{charbonneau02}. This observation foreshadowed a slew of spectra with truncated features \citep[e.g.][]{deming13, crossfield13, kreidberg14a, knutson14a, kreidberg15b}. Many spectra have also shown the large optical slope indicative of scattering from small particles \citep[e.g][]{lecavelier08b, sing11, sing13, robinson14, dragomir15}. Transit depth offsets between 1 and 5 $\mu$m are also seen in many hot Jupiters \citep{sing16}.

Despite this body of evidence, the makeup of the condensates has proven elusive. There are a number of theoretical possibilities, including equilibrium condensates such as water, salt, sulfide, or silicate clouds (depending on temperature) and photochemical hazes, e.g. hydrocarbon soots formed from photolyzed methane \citep{burrows99, kempton12, morley13, wakeford17}. Current transmission spectra lack the wavelength coverage and precision needed to distinguish between these species in exoplanets.  In brown dwarfs, a silicate feature at 9 $\mu$m has been tentatively detected using \Spitzer/IRS \citep{cushing06}. Future observations of exoplanets could also reveal features from specific grains, which would unambiguously determine their composition \citep{wakeford15}.

Meanwhile, there are several indirect constraints on condensate properties.  For example, optical phase curves of hot Jupiters are best explained by reflective clouds on their western hemispheres, composed of silicate or manganese sulfide \citep{demory13, oreshenko16, parmentier16}. Further clues come from the amplitude of spectral features. For example, the spectrum of the super-Earth GJ 1214b is featureless at high precision \citep[30 ppm,][]{kreidberg14a}.  Truncating the features to this extent requires an optically thick condensate layer at a pressure level of 0.1 millibar, which can be achieved either by thick, lofted clouds or very efficient haze formation \citep{morley15}. 

There is some evidence that condensates are more prevalent at lower temperatures \citep{stevenson16, heng16}. However, the microphysics of condensate formation are complex and depend sensitively on the thermal structure and circulation of the atmosphere \citep{turco79}. More atmosphere studies are needed to determine whether the presence of condensates can be predicted from temperature or other basic properties.

\section{Future Prospects}
We are on the threshold of a revolution in exoplanet atmosphere characterization thanks to several upcoming observing facilities.  The first of these is the Transiting Exoplanet Survey Satellite (\TESS), a planet finding mission scheduled to launch in 2018 \citep{ricker14}.  The goal of \TESS\ is to search 200,000 of the brightest stars in the sky for transiting planets, with an expected yield of nearly 2000 discoveries \citep{sullivan15}. In contrast to the majority of transiting planets discovered to date by \Kepler, the \TESS\ planets will have bright enough host stars for precise atmosphere characterization. The sample will also include smaller and cooler planets than are typically discovered by ground-based surveys of bright stars. 

The second game-changer is \JWST. The primary technical limitations in atmosphere characterization thus far have been aperture size and wavelength coverage, and \JWST\ offers major improvements on both fronts. It has roughly ten times the collecting area of \HST\, and provides spectroscopy from the optical to the infrared ($0.6 - 28\,\mu$m). The improvement in precision will make it possible to push down to sub-Neptune and smaller planets, and to study giant planets with unprecedented S/N. The expanded wavelength coverage will also enable the first spectroscopic detections of many molecules which have been elusive so far (including methane, carbon dioxide, and ammonia). Finally, \JWST's infrared detector MIRI will be sensitive to thermal emission from cooler objects, including potentially habitable worlds.

\runinhead{Pushing to Earth-Size Planets}
Atmosphere characterization for terrestrial planets is a challenge, but it is easiest for planets transiting nearby small stars.  Fortunately, such planets are common: roughly a quarter of M-dwarfs host a terrestrial planet in the habitable zone \citep{dressing15}. A number of these have been detected already, included three of the seven planets around the ultra-cool dwarf star TRAPPIST-1 and LHS 1140b \citep{gillon17, dittmann17}. Assuming these planets have high mean molecular weight atmospheric compositions like that of the Earth, the expected amplitude of spectral features in their atmospheres is of order 10 ppm -- a goal within reach of an intensive JWST observing campaign with repeated transit and eclipse observations for some of the planets \citep{morley17b}.  JWST is scheduled to launch in late 2018, so may soon provide us with the first ever constraints on terrestrial planet atmospheres beyond the Solar System.

\begin{acknowledgement}
The author acknowledges support from the Harvard Society of Fellows and the Harvard Astronomy Department Institute for Theory and Computation. She is grateful for helpful comments and figures from Caroline Morley, Hannah Diamond-Lowe, Tyler Robinson, and Sara Seager.
\end{acknowledgement}

\section{Bibliography}
\bibliographystyle{spbasicHBexo}  

\end{document}